\documentclass[manuscript]{aastex}

\usepackage{fancyhdr}
\usepackage{longtable}
\usepackage{latexsym}
\usepackage{graphicx}
\usepackage{amsmath}
\usepackage{natbib}
\usepackage{upgreek}
\usepackage{multirow}
\usepackage{array}
\usepackage{pdflscape}
\usepackage{graphicx,subfigure}
\usepackage{tabularx}
\usepackage{epsfig}
\usepackage{color}

\newcommand{\PreserveBackslash}[1]{\let\temp=\\#1\let\\=\temp}
\newcolumntype{C}[1]{>{\PreserveBackslash\centering}p{#1}}
\newcolumntype{R}[1]{>{\PreserveBackslash\raggedleft}p{#1}}
\newcolumntype{L}[1]{>{\PreserveBackslash\raggedright}p{#1}}

\newcommand       \mum          {\upmu\rm{m}}

\shorttitle{Spectra with features of crystalline silicates}
\shortauthors{Rui Chen et al.}

\begin{document}

\title{A systematic search for the spectra with features of crystalline silicates in the \textit{Spitzer} IRS Enhanced Products}

\author{Rui Chen\altaffilmark{1,2},
        Ali Luo\altaffilmark{1},
        Jiaming Liu\altaffilmark{3}, and
        Biwei Jiang\altaffilmark{3}}

\altaffiltext{1}{Key Laboratory of Optical Astronomy, National Astronomical Observatories, Chinese Academy of Sciences, Beijing 100012, China. \sf lal@nao.cas.cn}
\altaffiltext{2}{University of Chinese Academy of Sciences, Beijing 100049, China. }
\altaffiltext{3}{Department of Astronomy, Beijing Normal University, Beijing 100875, China}

\begin{abstract}
The crystalline silicates features are mainly reflected in infrared bands. The \textit{Spitzer} Infrared Spectrograph (IRS) collected numerous spectra of various objects and provided a big database to investigate crystalline silicates in a wide range of astronomical environments. We apply the manifold ranking algorithm to perform a systematic search for the spectra with crystalline silicates features in the \textit{Spitzer} IRS Enhanced Products available. In total, 868 spectra of 790 sources are found to show the features of crystalline silicate. These objects are cross-matched with the SIMBAD database as well as with the Large Sky Area Multi-Object Fibre Spectroscopic Telescope (LAMOST)/DR2. The average spectrum of young stellar objects show a variety of features dominated either by forsterite or enstatite or neither, while the average spectrum of evolved objects consistently present dominant features of forsterite in AGB, OH/IR, post-AGB and planetary nebulae. They are identified optically as early-type stars, evolved stars, galaxies and so on. In addition, the strength of spectral features in typical silicate complexes is calculated. The results are available through CDS for the astronomical community to further study crystalline silicate.
\end{abstract}


\section{Introduction}

In the interstellar space silicates get modified, destroyed and potentially reformed, they play an important role in the cosmic life cycle of matter. As an important component of circumstellar and interstellar dust, silicates provide information about the size of grains and the chemical composition of dust. Their mid-infrared spectral features can help us to get the information about the density and thermal structure of circumstellar disks and envelopes \citep[see][]{Henning2010}.

All the earlier observations showed that the feature bands of silicates are broad and smooth around 9.7$\mu$m and 18$\mu$m \citep[see][]{Molster&Kemper2005} respectively.  \citet{Dorschner et al.1995} used laboratory data to identify the carrier of these features to be amorphous olivine silicates. Later, the Infrared Space Observatory (ISO) and the \textit{Spitzer} Space Telescope obtained a lot of infrared spectra with high sensitivity and high resolution, which opened a new window for us to identify and investigate the spectral features of crystalline silicates.  \citet{Molster et al.2002a,Molster et al.2002b,Molster et al.2002c} systematically studied the spectral features of crystalline silicate and diversified them into 7 distinct complexes  using the ISO/SWS (Short Wavelength Spectrometer) and LWS (Long Wavelength Spectrometer) spectra, which are called the 10$\mum$ (7-13$\mum$), 18$\mum$ (15-20$\mum$,), 23$\mum$ (22-25.5$\mum$,), 28$\mum$ (26.5-31.5$\mum$), 33$\mum$ (31.5-37.5$\mum$), 40$\mum$ (38-44.5$\mum$) and 69$\mum$ (50-72$\mum$) complex respectively.

Crystalline silicates have been identified in various scales (solar system, stars, galaxies and quasars), such as comet, protoplanetary disk of Herbig Ae/Be stars and T Tauri stars, debris disks around the main sequence stars \citep{Li&Greenberg1998}, circumstellar envelopes in post-main sequence stars \citep{Water et al.1996,Sylvester et al.1999,Molster et al.2002a,Jiang et al.2013}, very bright infrared galaxies \citep{Markwick-Kemper et al.2007,Spoon et al.2006}, quasars \citep{Markwick-Kemper et al.2007,Aller et al.2012,Xie et al.2014}, etc.

In this work, we present a systematic search for the spectra with features of crystalline silicates
from the \textit{Spitzer} IRS Enhanced Products \citep{Houck et al.2004}. In Section~\ref{sec:Dd}, a brief description is presented about the \textit{Spitzer} IRS Enhanced Products \citep{Houck et al.2004}. The method we used is manifold ranking (MR) algorithm firstly proposed by \citet{Zhou et al.2004a,Zhou et al.2004b}, which will be described in Section~\ref{sec:MR}. 
The selected objects with crystalline silicate features are cross identified with the SIMBAD database as well as with the LAMOST optical spectroscopy in Section~\ref{sec:Id}.

\section{Data description}
\label{sec:Dd}
\begin{figure}
\centering
\includegraphics[height=7.5in,width=\textwidth]{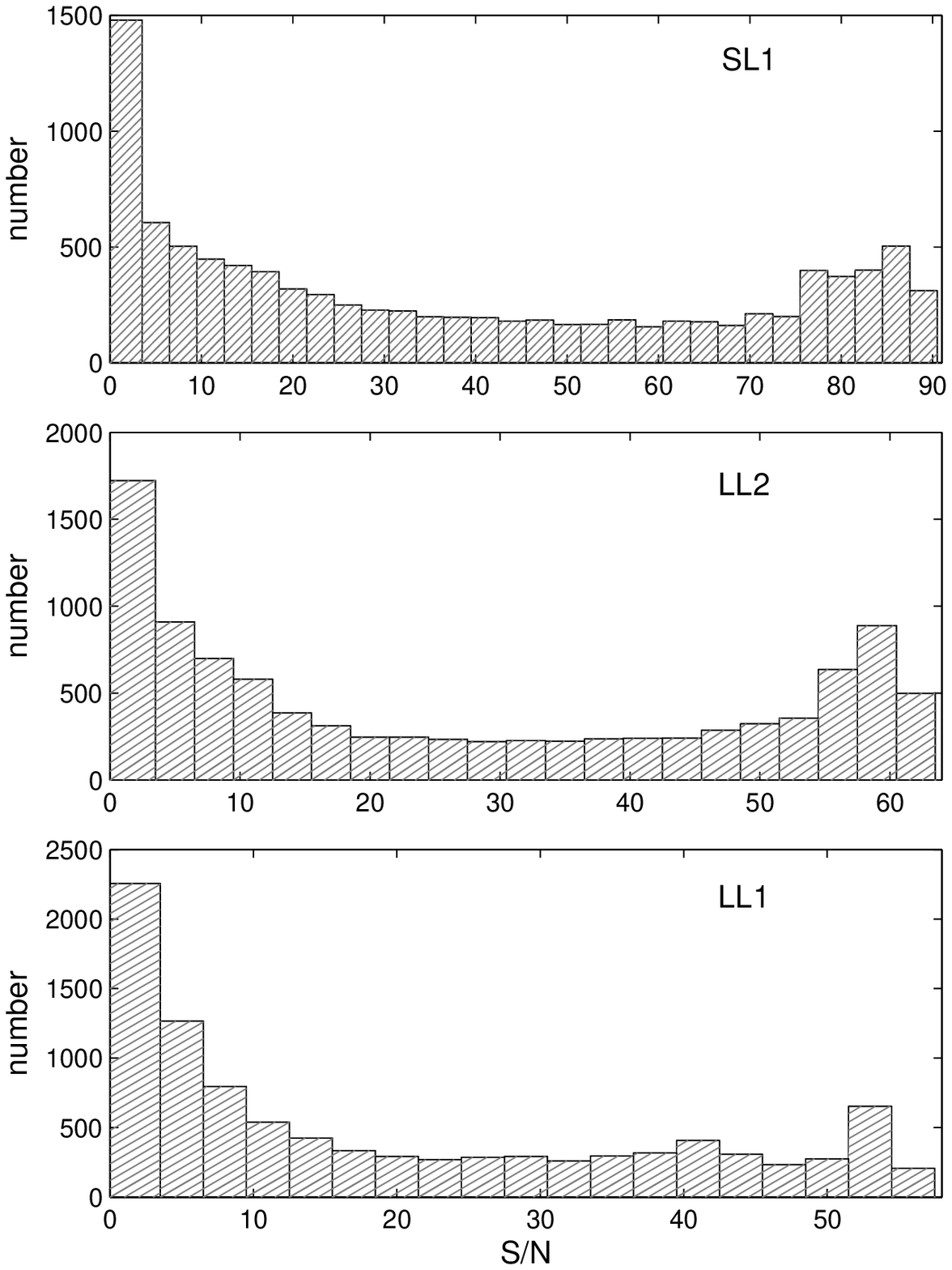}

\caption{S/N distribution of the basic 9711 Spitzer/IRS spectra.}
\label{fig:1}
\end{figure}

\begin{figure}
\centering
\includegraphics[height=5.0in,width=\textwidth]{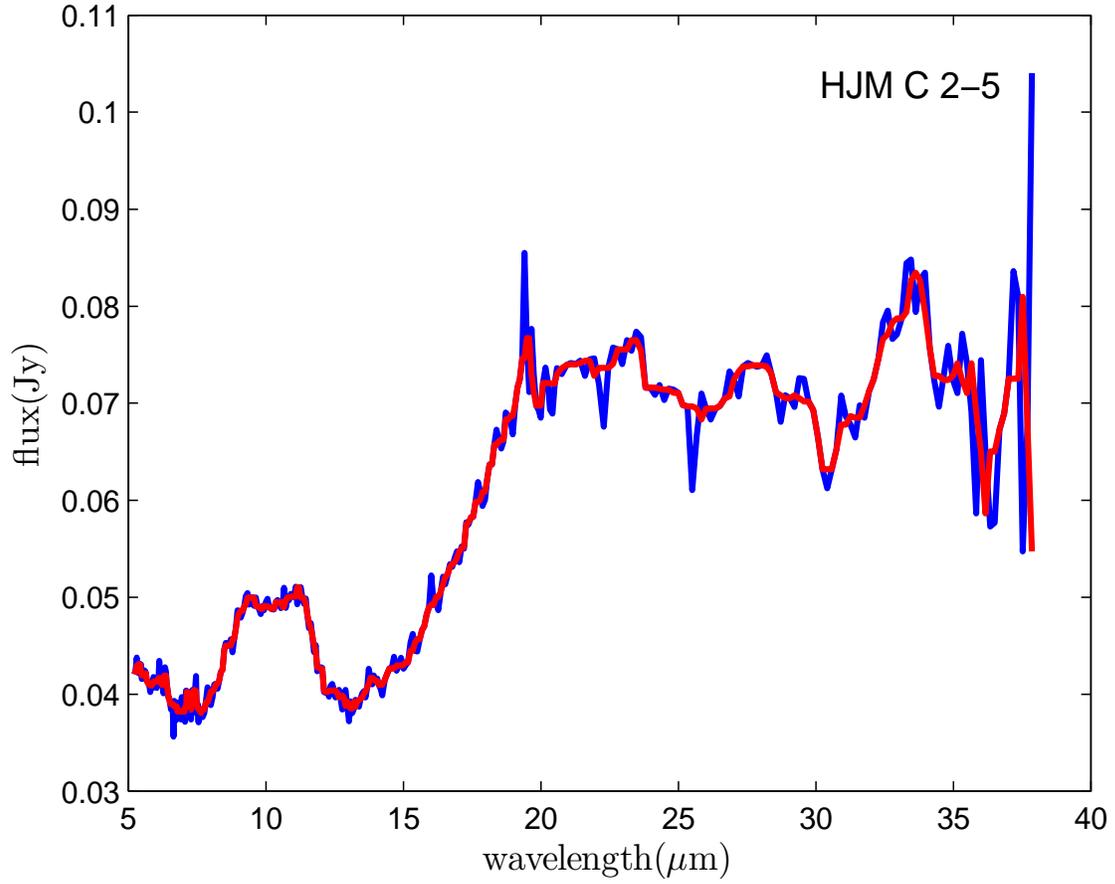}

\caption{The effect of median filtering. The blue spectrum is the observed spectrum, while the red one is the processed spectrum after median filtering.}
\label{fig:3}
\end{figure}

Spectra taken by the InfraRed Spectrograph (IRS) \citep{Houck et al.2004} on the \textit{Spitzer} space telescope \citep{Werner et al.2004} are now publicly available. These spectra are produced using the bksub.tbl products from SL and LL modules of final SSC pipeline, version 18.18. From the IRS data archive, we found a collection of 16986 low-resolution spectra. The spectra are merged by 4 slits which are SL2 (5.21$\sim$7.56$\mu$m), SL1 (7.57$\sim$14.28$\mu$m), LL2 (14.29$\sim$20.66$\mu$m) and LL1 (20.67$\sim$38.00$\mu$m). As crystalline silicates have no features in the SL2 band, we choose the spectra which include all the other three bands SL1, LL2 and LL1 so that the object has a continuous spectrum from about 7.5$\mum$ to 38$\mum$. In this way, five of the seven infrared complexes of crystalline silicates are covered, i.e. the 10, 18, 23, 28 and 33$\mum$ complexes. Consequently, 9711 spectra are picked up. Fig.~\ref{fig:1} shows the S/N distribution of these 9711 spectra for the three IRS slits respectively. In order to reduce the noise, median filtering is applied. The main idea of median filtering is to run through the spectrum pixel by pixel, replacing each pixel with the median of neighboring pixels. Here we use median filtering with a width of 5 pixels which means each pixel is replaced with the median of 5 neighboring pixels. Because the spectral features of silicate dust, even of crystalline silicate, are much wider than the atomic and ionic lines, they are not affected by the median filtering. Fig.~\ref{fig:3} compares the spectrum before and after the median filtering and proves its efficiency in depressing the high-frequency noise and spikes.

\section{Search for spectra with crystalline silicates features}

\subsection{Method and result}

\label{sec:MR}

Considering a sample set $\mathcal{X}=\{x_{1},\cdots,x_{q},x_{q+1},\cdots,x_{n}\}$, the first $q$ units are labelled and belong to the same class (here are the known spectra with crystalline silicate features), and the rest are unlabelled (here are 9711 spectra from the \textit{Spitzer} IRS). Our task is to retrieve the units similar to the labelled ones. In general, the labelled sample is much smaller compared with the whole sample, namely $q<<n$. Manifold ranking (MR) \citep{Zhou et al.2004a, Zhou et al.2004b} can rank the unlabelled units according to their similarity to the labelled and improve the results by using the relationship between unlabelled and labelled units.

Let $F:\mathcal{X} \longrightarrow R$ denotes a ranking function which assigns to each unit $x_{i}$ a ranking value $F_{i}$. Let $Y = [Y_{1},\ldots,Y_{n}]^{T}$, in which $Y_{i}=1$ if $x_{i}$ is a labelled unit, and $Y_{i}=0$ otherwise. The procedure of MR is as follows:
\begin{enumerate}
\item Construct the weighted graph $G=(X,W)$ by using a kNN (k Nearest Neighbor) graph, where $W$ is a symmetric adjacency matrix. $W_{ij}$ represents similarity between unit $x_{i}$ and $x_{j}$  and is defined by:
    \begin{equation}
        W_{ij}=
        \begin{cases}
        e^{\frac{-d^{2}(x_{i},x_{j})}{2\sigma^{2}}} & \text{$x_{i}$,$x_{j}$ are connected and $i\neq j$}\\
        0 & \text{others}
        \end{cases}
    \end{equation}
    where $d(x_{i},x_{j})$ is the distance between $x_{i}$ and $x_{j}$, here we use the Euclidean distance, and $\sigma$ is a constant that controls the strength of the weight.
\item Form the matrix $S = D^{-1/2}WD^{-1/2}$, where $D$ is a diagonal matrix with its (i,i)-element equals to the sum of the i-th row of W.
\item Iterate $F(t+1) =\alpha SF(t)+(1-\alpha)Y$ until convergence, where $\alpha$ is a parameter in (0,1). The intuitive description of this step is that all the units spread their scores to the nearby units via the weighted graph.
\item Let $F^{*} = F(t)$, sort $F^{*}$ in descending order and return index rank.
\end{enumerate}

\begin{figure}
\centering
\includegraphics[height=5.0in,width=\textwidth]{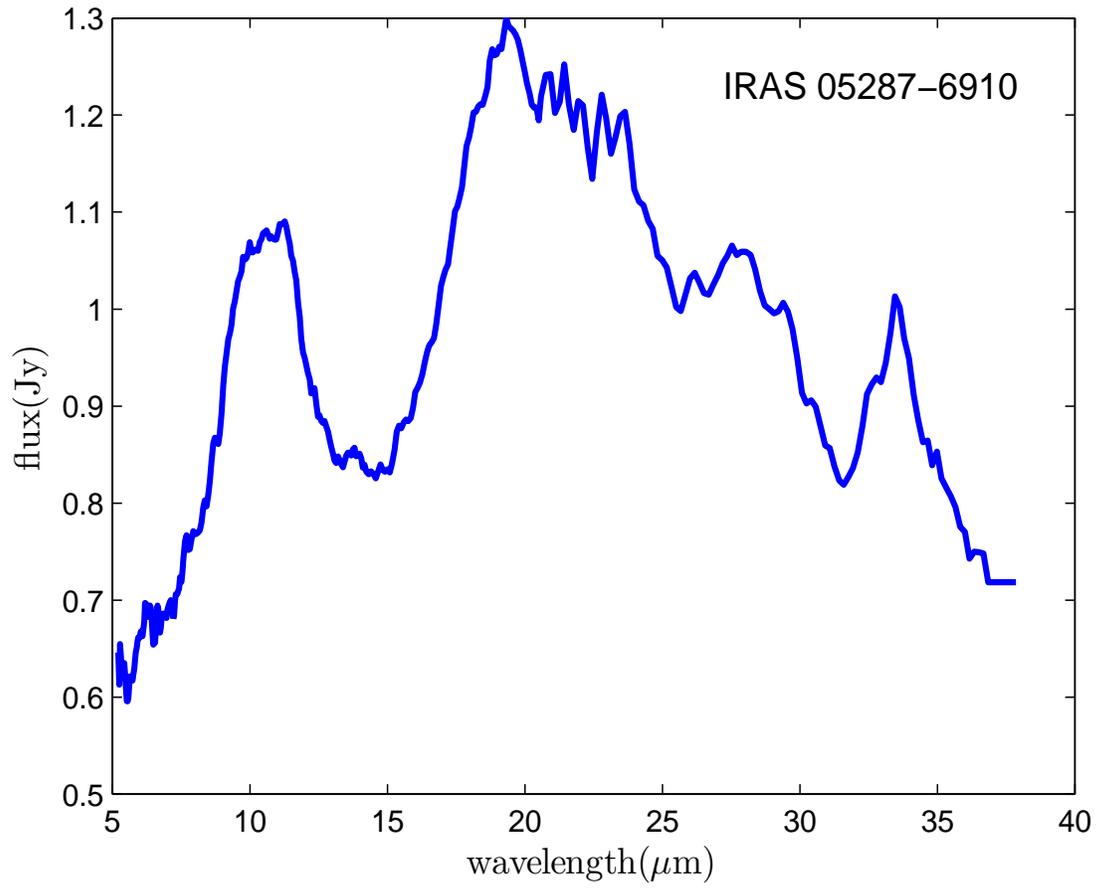}

\caption{An example spectrum with crystalline silicates in \citet{Jones et al.2012}.}
\label{fig:2}
\end{figure}

The basis of the labelled sample for MR is the 38 Spitzer/IRS spectra that have crystalline silicate features from \citet{Jones et al.2012}, which is a relatively large sample and also taken by \textit{Spitzer}/IRS. Our mission is to retrieve the spectrum which has crystalline silicate features similar to the labelled samples. It should be noted that 28 of the 38 Jones's sources are within the 9711 spectra, which means some duplication. As an example, Fig.~\ref{fig:2} shows one of the spectra, with prominent emission at the complexes of crystalline silicates. Before the MR algorithm is performed, the median-filtered spectrum is normalized by Minimum-Maximum standardization using following formula:
\begin{equation}
sp_{i}=\frac{sp_{i}-min(\textbf{\textit{sp}})}{max(\textbf{\textit{sp}})-min(\textbf{\textit{sp}})}
\end{equation}
where $sp_{i}$ is the i-th pixel of the spectrum $\textbf{\textit{sp}}$, so that the maximum is normalized to unique for the convenience of calculation.

In order to verify the completeness and stability of MR, we test the result by varying the number of labelled units. We randomly select a given number $n$ of spectra with crystalline silicates features from the 38 sources to compose the labelled sample, while the non-selected spectra with crystalline silicates features and all the 9711 spectra from \textit{Spitzer}/IRS consist  the unlabelled sample.  We vary the number $n$ of labelled units in \{1, 10, 19, 28, 37\}. Then we use MR to get the ranking list. We use $n$ to denote the number of labelled units, $R@k$ to denote what fraction of the 38 known spectra with crystalline silicates features are retrieved within the top $k$ ranking spectra and $NI@k$ to denote $m/k$ where $m$ is the number of intersection within the top $k$ ranking spectra between different $n$.

\begin{figure}
\centering
\subfigure[] {\includegraphics[height=3.3in,width=\textwidth]{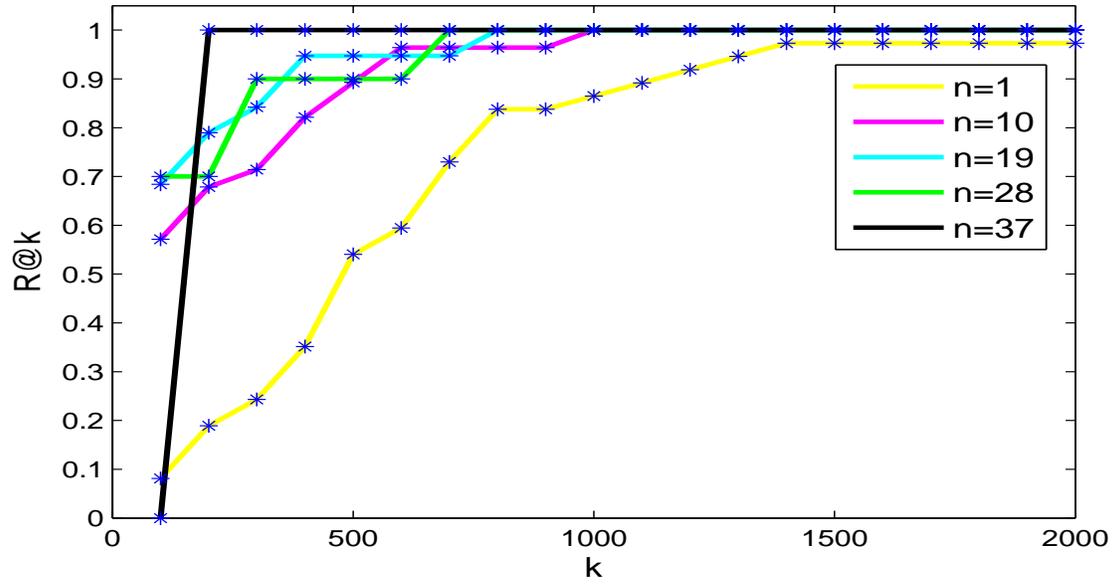}}
\hskip 0.03in
\subfigure[] {\includegraphics[height=3.3in,width=\textwidth]{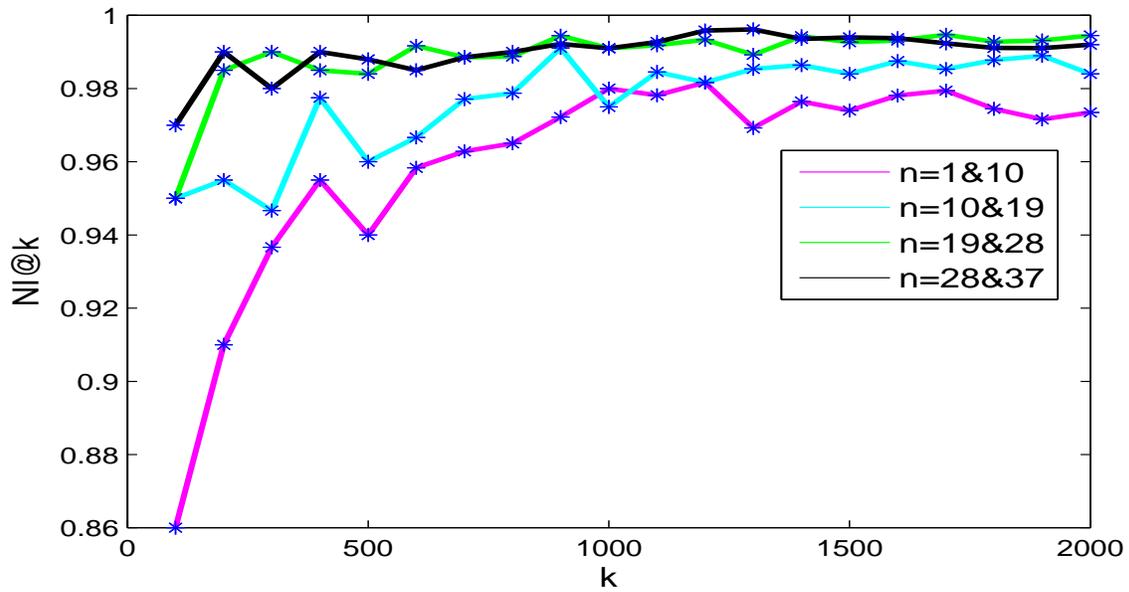}}

\caption{Upper panel:  comparison of R@k for different settings of $k$. Lower panel: comparison of NI@k. The legend of right figure, for example, $n$=1\&10 means comparison between the results of $n$=1 and $n$=10.}
\label{fig:suanfa}
\end{figure}

The test results of different $n$ are displayed in Fig.~\ref{fig:suanfa}. From the left figure, it can be seen that the non-selected spectra with crystalline silicates' features are found within top 900 ranking when $n$=10, 19, 28 and 37, while 97\% non-selected spectra are found within top 1300 ranking when $n$=1. Meanwhile, the right figure tells that the ranking list is stable against the change of the number of labelled samples when $n\geq$19.

\begin{figure}
\centering
\includegraphics[height=7.5in,width=5.8in]{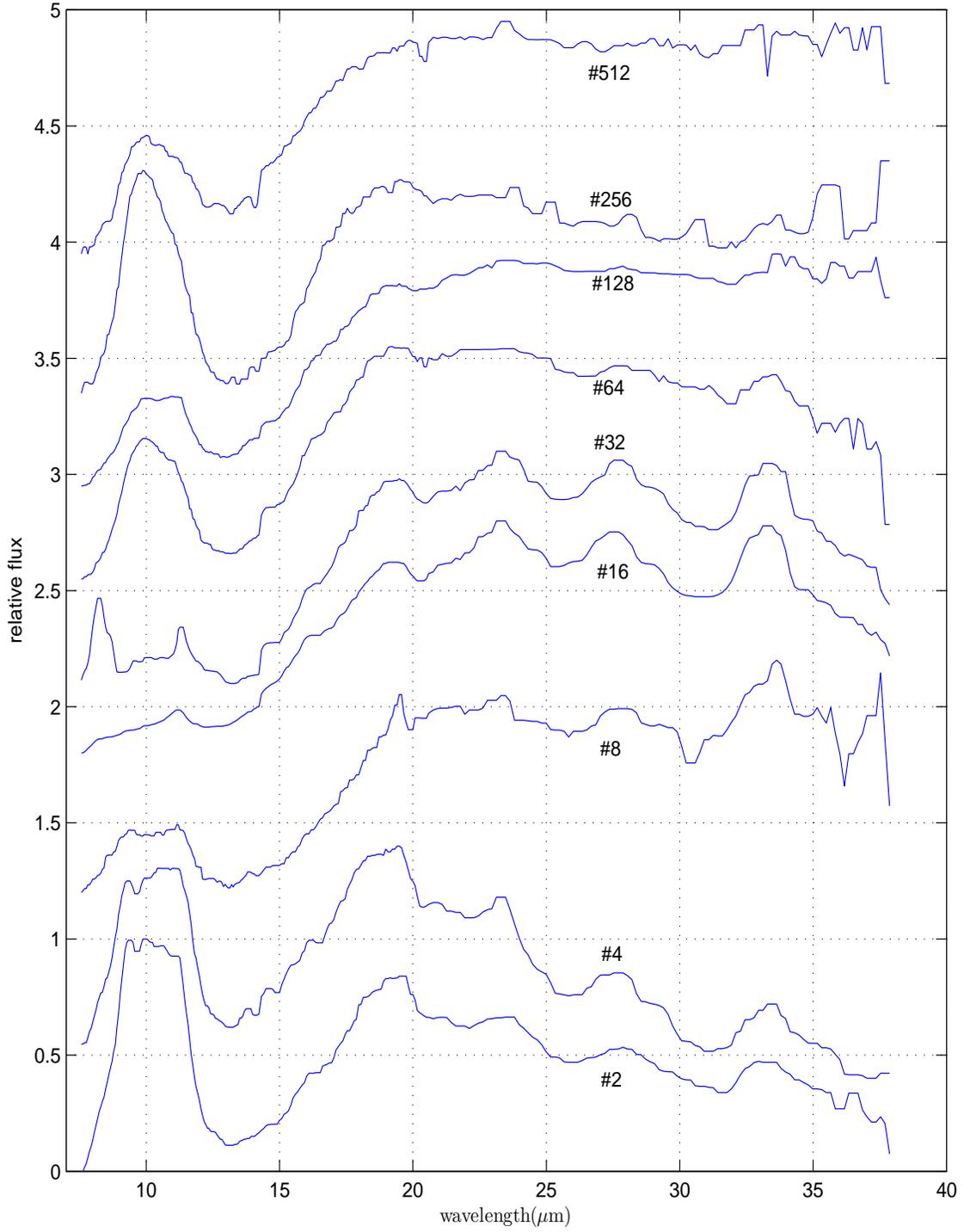}

\caption{The spectra of sources found by manifold ranking. The ranking numbers are  2, 4, 8, 16, 32, 64, 128, 256 and 512 respectively.}
\label{fig:24816}
\end{figure}

\begin{figure}
\centering
\includegraphics[height=7.5in,width=\textwidth]{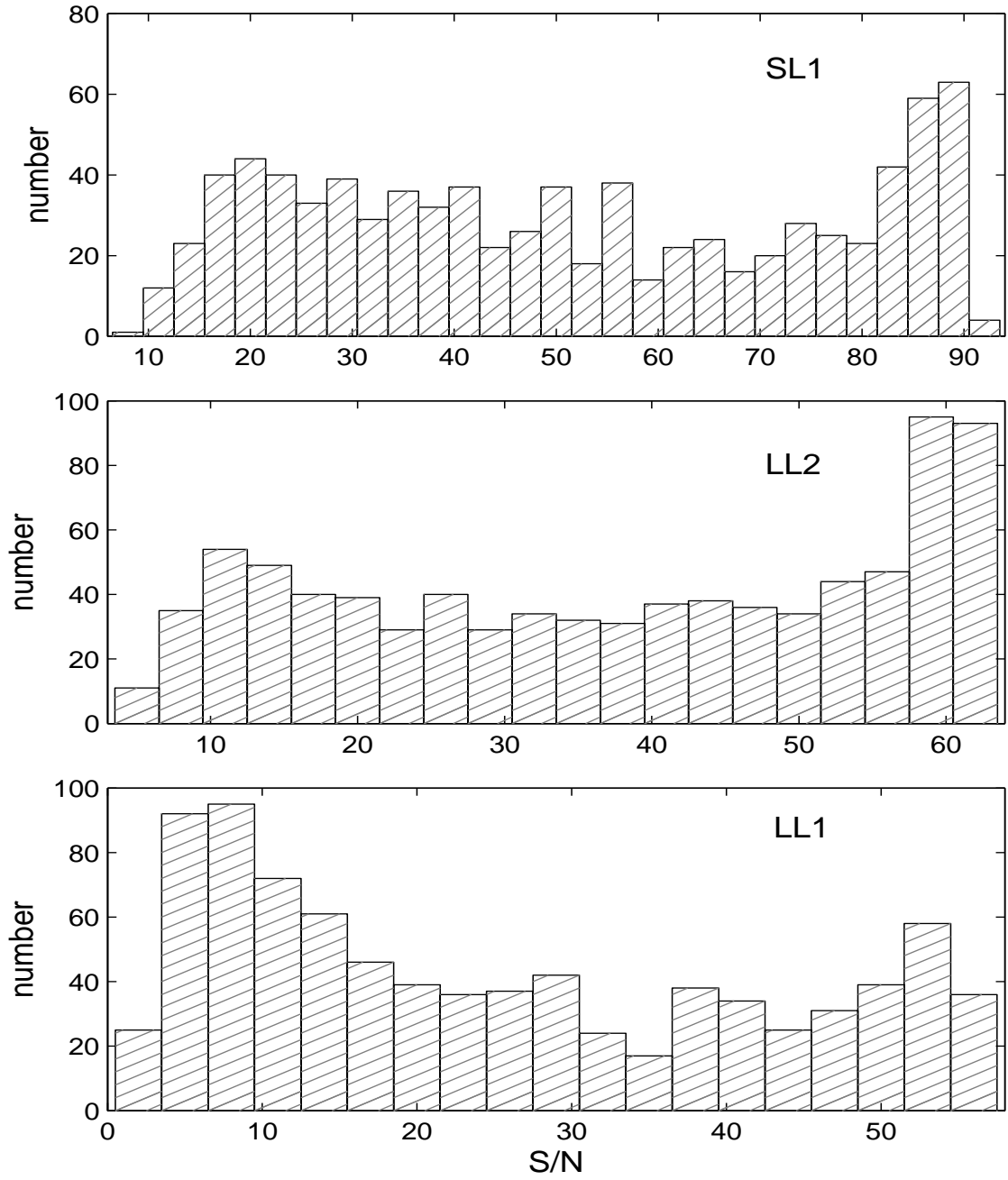}

\caption{S/N distribution of the 868 spectra we find.}
\label{fig:snr868}
\end{figure}

The top 2000 ranking samples are selected for final eye-checking. 868 spectra are found to show crystalline silicates features (see Table~\ref{tab:MR}). From Table~\ref{tab:MR} we can see the spectrum has higher probability with crystalline silicates features when its ranking is higher, and we can hardly find any spectra with crystalline silicates features when their rankings are behind 2000. The spectra of the sources ranking at 2, 4, 8, 16, 32, 64, 128, 256 and 512 are displayed in Fig.~\ref{fig:24816}. The S/N distribution of these 868 spectra in Fig.~\ref{fig:snr868} indicates that most of the sources have $S/N > 10$ in the SL1 and LL2 bands ($\sim7-21\mum$), much better than the average of primordial 9711 sample objects.

\subsection{Intensity of the spectral features of crystalline silicates}

\begin{figure}
\centering
\includegraphics[height=4.0in,width=\textwidth]{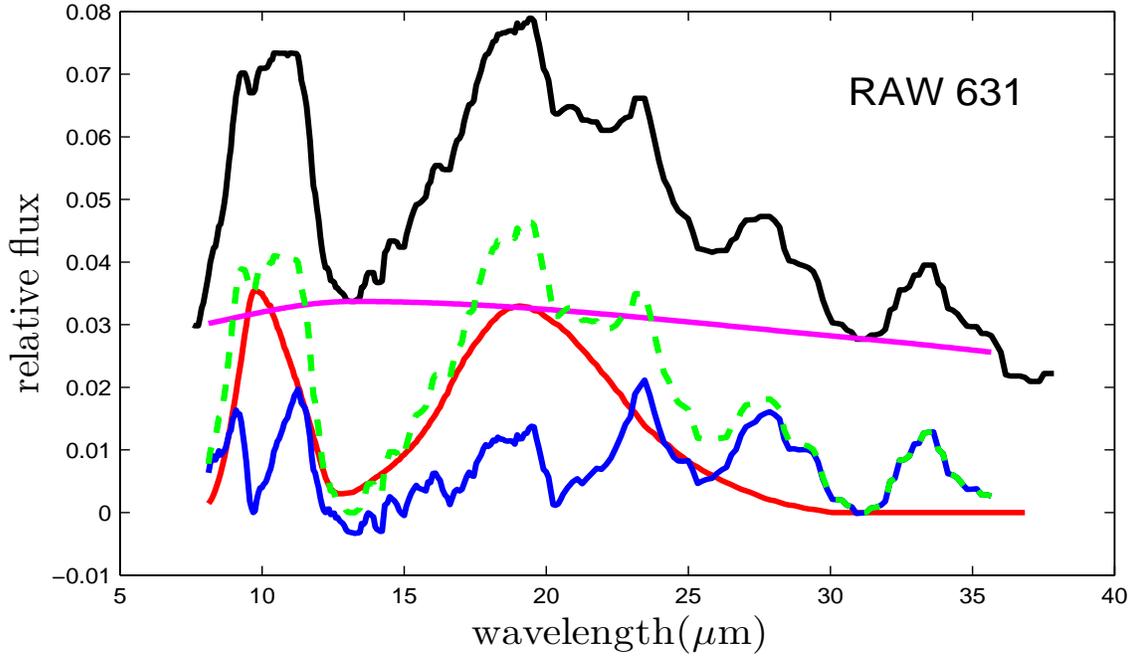}
\caption{Fitting of the continuum. Black line is the observed spectrum,  pink line is the dust continuum which is fitted by a polynomial function based on the anchor points described in the text, green dotted line is the observed spectrum after the dust continuum subtracted which is decomposed into two parts  -- the amorphous silicate features (red line) and the crystalline silicate features (blue line).}
\label{fig:con}
\end{figure}

We consider that the observed flux is contributed by dust continuum ($F_{\rm cont}$), and dust spectral features from both amorphous silicate  ($F_{\rm amo}$) and crystalline silicate ($F_{\rm crys}$):
\begin{equation}
F_{\rm obs} = F_{\rm cont}+F_{\rm amo}+F_{\rm crys}
\end{equation}
For each spectrum, we first modeled the dust continuum by fitting a piecewise polynomial of degree 3 to anchor points. Following the analysis of \citet{Watson et al.2009}, we selected the following anchor points: 7.57-7.94, 13.02-13.50, 30.16-32.19, and 35.07-35.92$\mum$ which are not affected by solid-state emission features.  After the continuum is subtracted, the amorphous silicate features around 10$\mum$ and 18$\mum$ are fitted by:
\begin{equation}
F_{\rm amo}(\nu) \sim B(T_{\rm amo},\nu) \times Q_{\rm amo}(\nu)
\end{equation}
where $B(T_{\rm amo},\nu)$ is a blackbody radiation at the temperature $T_{\rm amo}$, and $Q_{\rm amo}(\nu)$ is the absorption efficiency of amorphous silicate. The absorption efficiency of amorphous silicate is based on the optical constants of \citet{Ossenkopf et al.1992} assuming the dust to be spherical with radii $a=0.1\mum$, following the numerals of \citet{Jiang et al.2013}. The temperature of  amorphous silicate is adjusted in the range of 40-1200\,K at an interval of 10\,K to achieve the best fitting. The residual after subtracting the dust continuum and the spectral features of amorphous silicate is taken to be the spectral features of crystalline silicate. Fig.~\ref{fig:con} shows an example. It can be seen that the amorphous silicate features take a priority over the crystalline silicate since fitting the amorphous spectral features are maximized to match the spectrum after the continuum is subtracted. This fitting may under-estimate the intensity of the spectral features of crystalline silicate, which means we are obtaining a lower limit of the intensity ratio of crystalline to amorphous silicate's features in the following.

The intensity of crystalline silicate bands is measured in order to understand their properties in relation to stellar parameters. Here the equivalent widths of the 10, 18, 23, 28 and 33$\mum$ features, $W_{\rm eq}$ is defined as:
\begin{equation}
W_{\rm eq} = \sum_{\lambda}\frac{F_{\rm crys}}{F_{\rm cont}}\triangle\lambda
\end{equation}
The wavelength interval of each feature agrees with that defined by \citet{Molster et al.2002a} and described in Section 1.


The significance level of the crystalline silicate is calculated in each band as
\begin{equation}
S/U = \frac{F_{\rm crys}}{\sigma_{\rm crys}}
\end{equation}
where $F_{\rm crys}$ is crystalline silicate flux and $\sigma_{\rm crys}$ is the uncertainty in the measured flux,
\begin{equation}
F_{\rm crys} = \sum_{\lambda} (F_{\rm obs}-F_{\rm cont}-F_{\rm amo})\Delta\lambda
\end{equation}
\begin{equation}
\sigma_{\rm crys}^{2} = \sum_{\lambda}((X_{\rm fac}*(F_{\rm cont}+F_{\rm amo}))^{2}+\sigma^{2}) \Delta\lambda^{2}
\end{equation}
where $\sigma$ is statistical uncertainty in each resolved element of
the spectra, and $X_{\rm fac}$ is the uncertainty in the determination of
the underlying continuum and amorphous silicate, which we conservatively assumed
to be 10\% ($X_{\rm fac} = 0.1$). \textbf{The value of $X_{\rm fac}$ is the same as that adopted by \citet{Mittal et al.2015} since we both deal with the \emph{Spitzer}/IRS spectrum and determine the continuum in a similar way. In addition, some experiments are carried out by choosing the neighbouring points of default anchor points for the determination of continuum and amorphous silicate's features. It is found the scattering of the dust continuum and amorphous silicate features are both less than 0.1.} The results of significance level in each band are listed in Table~\ref{tab:significanceLevel}.

We also calculate the crystallinity which can provide a quantitative measure of the percentage of crystalline silicate and defined as
\begin{equation}
Crystallinity = \frac{\sum_{\lambda} F_{\rm crys} \Delta\lambda}{\sum_{\lambda} F_{\rm crys} \Delta\lambda+\sum_{\lambda} F_{\rm amo} \Delta\lambda}
\end{equation}
The results of crystallinity are listed in Table~\ref{tab:significanceLevel}. Considering that amorphous silicate can contribute to the continuum emission, a lower limit of crystallinity defined by flux ratio is also calculated as:
\begin{equation}
Crystallinity\_min = \frac{\sum_{\lambda} F_{\rm crys} \Delta\lambda}{\sum_{\lambda} F_{\rm obs} \Delta\lambda}
\end{equation}
The results are in Table~\ref{tab:significanceLevel}.

\subsection{Identification in the SIMBAD and LAMOST/DR2 databases}
\label{sec:Id}

The 868 spectra are cross identified with the SIMBAD database to learn the types of sources and the results are presented in Table~\ref{tab:2}. Removing duplicate sources, there are 790 sources whose classification is summarized  in Table~\ref{tab:3}. Many of the sources are in the early stage of evolution, including the Orion-type variables, T Tauri stars, pre-main sequence stars, Herbig Ae/Be stars, FU Ori type. The second most popular class is in the late stage of evolution. They are mostly the low-mass evolved stars such as AGB stars, OH/IR stars, post-AGB stars or proto-planetary nebulae, and planetary nebulae, as well as carbon stars and S-type stars. At the same time, the massive stars in late stage of evolution are also present, such as red and blue supergiants. The detection of crystalline silicate has been reported widely in the early and late stage of stellar evolution. In addition, crystalline silicate emission is found in different types of variables, such as nova, BY Dra type, RR Lyr type, delta Sct type, and several types of long-period pulsating variables. Many types have only one object in our identified sample of crystalline silicate emitters. Confirmation of these types of objects needs more serious investigation.

\begin{figure}
\centering
\includegraphics[height=7.0in,width=\textwidth]{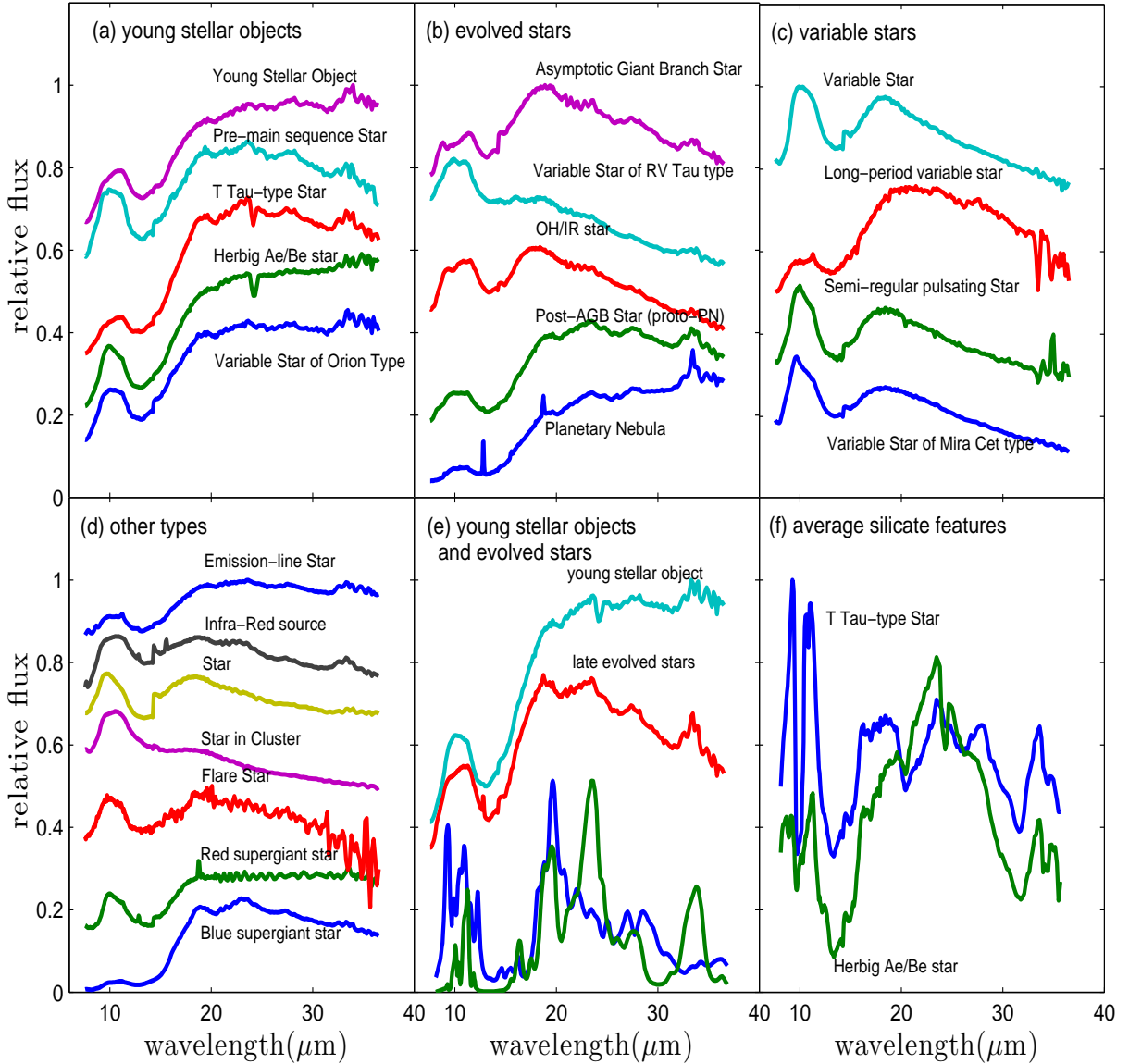}
\caption{The averaged spectrum for each type with $\geq$ 5 objects. In the (e) figure, the average of all young stellar objects and evolved stars are compared with the mass absorption coefficients \citep{Koike et al.1999} of forsterite (red line) and enstatite (green line). \textbf{In the (f) figure, the spectral features of silicates after subtracting the continuum are compared between T Tau-type and Herbig Ae/Be stars.} }
\label{fig:avefig}
\end{figure}

The average spectrum is calculated for the types which have more than five objects by a weighted mean according to the flux intensity. It should be noted that the objects can be generally divided into two major classes, i.e. the young stellar objects and the evolved stars, even though the SIMBAD types are various. The young stellar objects may include all the types in the early stage of evolution, such as pre-main sequence star and proto-star, while pre-main sequence star may include the low-mass T Tau-type star (which can also be Orion-type variable) and massive Herbig Ae/Be star. On the other hand, AGB star, OH/IR star, post-AGB star, planetary nebula etc. are all in the late evolutionary stage of low-mass star. The variable stars (semi-regular, long-period, Mira-type) are overlapped with the evolved stars. As shown in Fig.~\ref{fig:avefig}, we divided the objects into four major classes: young stellar objects, evolved stars, variable stars and other types. In the final, the average spectra of all young stellar objects and all evolved stars are compared in Fig.~\ref{fig:avefig} (e). It can be seen that the young stellar objects generally have colder continuum than evolved stars. In respect to the spectral features of crystalline silicate, the evolved stars have wider feature around 10$\mum$, and show a more evident peak at about 11.2$\mum$. In comparison with the absorption coefficient of typical crystalline silicates -- forsterite and enstatite, such profile points to a richer amount of forsterite in the evolved stars than in young stellar objects. Consistently, the features at 20$\mum$ and 24$\mum$ of forsterite are also prominent. Nevertheless, it should be noted that the young stellar objects exhibit diversity among sub-types:  T Tau-type stars have a raised red wing like evolved stars, Herbig Ae/Be star have a raised blue wing, while the Orion-type variables show a flat top around 10$\mum$. Such distinction between T Tau-type stars and Herbig Ae/Be stars is different from previous studies. It was believed that the spectra of T Tau-type stars in the 10$\mum$ range trace the warm inner disk and thus show more enstatite features \citep[e.g.][]{Bouwman et al.2008} while Herbig Ae/Be stars trace the cold outer disk and thus more forsterite features \citep[e.g.][]{Juhasz et al.2010}. Our average spectra show a contrary tendency. \textbf{A possible reason for this discrepancy is that the average spectrum may be affected by the dust continuum. Removing the dust continuum emission, we can see that the T Tau-type stars show the features dominated by enstatite and the Herbig Ae/Be stars show the features dominated by forsterite in the 10$\mum$ range as can be seen in Fig.~\ref{fig:avefig} (f). This is then consistent with previous conclusion.} On the other hand, the evolved stars conform to the average spectrum. AGB stars, OH/IR stars, post-AGB stars and planetary nebulae all point to a raised red wing of the 10$\mum$ feature, indicating the dominant role of forsterite. This consistent scenario implies similar condition for crystallization of silicate dust in the circumstellar envelope of evolved stars. One exception is the RV Tau-type variables which shows no dominance of spectral features of forsterite. As RV Tau-type stars are usually post-AGB stars in a binary system, there may be disk-like structure to influence the composition of crystalline silicate.

\begin{figure}
\centering
\includegraphics[height=3.5in,width=\textwidth]{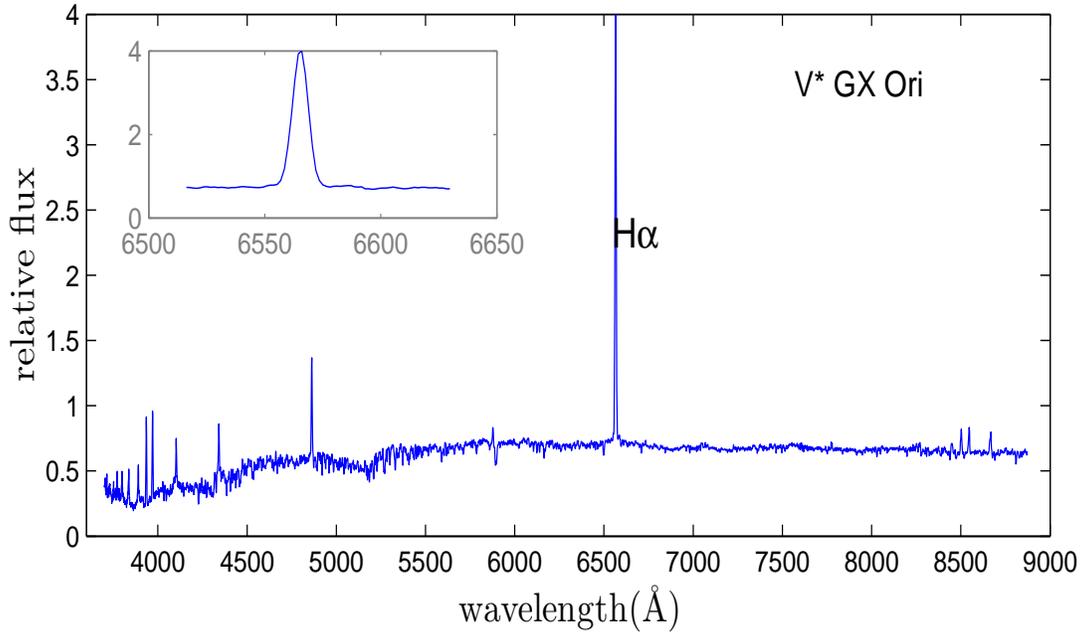}

\caption{One LAMOST optical spectra with H$\alpha$ emission line. The H$\alpha$ line is shown in detail in the inset.}
\label{fig:5}
\end{figure}

The identified sample is also cross matched with the LAMOST/DR2 catalog resulting in 91 identifications. As LAMOST is an optical spectroscopic survey, the objects are classified spectroscopically, with their types listed in Table~\ref{tab:4} and the summary of spectral types in Table~\ref{tab:5}. Very interestingly, 83 of 91 (i.e. 91\%) sources have strong H$\alpha$ emission line. The H$\alpha$ emission is usually associated with circumstellar disk or envelope in early-type stars or chromospheric activity in late-type dwarf stars. Whether the condition to excite the H$\alpha$ emission is related to the crystallization of silicate deserves further investigation. From this high percentage of H$\alpha$ emission occurrence in the LAMOST identified sample, there should be some connection between these two phenomena.  The types span the stellar classes M-, K-, G-, F- and A-type, and one white dwarf and one quasar. A careful check confirmed the spectral classification, except the white dwarf identification. On the other hand, 3 stars have strong H$\alpha$ absorption line. Fig.~\ref{fig:5} shows one optical LAMOST spectrum with the H$\alpha$ emission line.

%

\section{Summary}
We performed a systematic search for objects with spectral features of crystalline silicate in the \emph{Spitzer}/IRS enhanced products using the manifold ranking method. The primary results of our study are as follows.
\begin{enumerate}
  \item We identified 790 sources whose IRS spectrum presents the crystalline silicates' features.
  \item The equivalent widths of the 10, 18, 23, 28 and 33$\mu$m features of each spectrum are calculated to characterize the strengths of crystalline silicates features.
  \item The sources with crystalline silicates are cross-identified  in the SIMBAD database for the type of each object. The average spectrum is calculated for the types with more than five objects. It is found that the average spectrum of young stellar objects shows a variety with the features dominated either by enstatite or by forsterite, while the evolved stars all show dominance of the spectral features of forsterite in AGB, OH/IR stars, post-AGB stars and planetary nebulae.
  \item The crystalline silicate stars are cross-identified with the LAMOST spectroscopy which results in 91 optically identified objects. It is found that 83 (91\%)of them show H$\alpha$ emission line.
\end{enumerate}

\section*{Acknowledgements}

We thank Dr. Ke Zhang for her help in fitting the spectra and the anonymous referee for very helpful suggestions. This work was based on observations made with the \textit{Spitzer} Space Telescope, which is operated by the Jet Propulsion Laboratory, California Institute of Technology under a contract with NASA. The work was funded by the National Science Foundation of China (Grant Nos. 11373015, 11390371, 11533002) and the National Basic Research Program of China (973 Program, 2014CB845700). The Guo ShouJing Telescope (the Large Sky Area Multi-Object Fiber Spectroscopic Telescope, LAMOST) is a National Major Scientific Project built by the Chinese Academy of Sciences. Funding for the project has been provided by the National Development and Reform Commission. LAMOST is operated and managed by the National Astronomical Observatories, Chinese Academy of Sciences.

\clearpage

\clearpage

%
%
%
%

\clearpage

\begin{table}
\centering

  \caption{The results of manifold-ranking.}
 \label{tab:MR}%
    \begin{tabular}{@{}lccccccccccl}
   \tableline\tableline

    ranking range & N$_{\rm crystalline}$ & ranking range & N$_{\rm crystalline}$  \\

     \tableline
    1$\sim$100 & 98 & 1001$\sim$1100 & 41 &  \\
    101$\sim$200& 91 & 1101$\sim$1200 & 30&   \\
    201$\sim$300 & 80 & 1201$\sim$1300 & 22 &  \\
    301$\sim$400 & 78 & 1301$\sim$1400 & 13 &  \\
    401$\sim$500 & 76 & 1401$\sim$1500 & 7 &   \\
    501$\sim$600 & 74 & 1501$\sim$1600 & 4 &   \\
    601$\sim$700 & 70 & 1601$\sim$1700 & 2 &  \\
    701$\sim$800 & 74& 1701$\sim$1800 & 1 &  \\
    801$\sim$900 & 60& 1801$\sim$1900 & 0 &  \\
    901$\sim$1000 & 50 & 1901$\sim$2000 & 1 &  \\

    \tableline
    \end{tabular}%
 \tablenotetext{a}{ The column of N$_{\rm crystalline}$ is the number of spectra with crystalline silicates features.}
\end{table}%

\clearpage

\begin{table}
\centering
  \caption{Wavelength intervals for extracting continuum.}
\label{tab:int}

\begin{tabular}[]{cccc}
\tableline\tableline
feature($\mu$m) & continuum interval($\mu$m) \\
\tableline
10 & 9.5-11.5 \\
18 & 17.0-20.0 \\
23 & 22.3-24.6 \\
28 & 26.0-31.1 \\
33 & 31.5-35.0 \\
\tableline
    \end{tabular}%
\end{table}%

\clearpage

\begin{table}
\setlength\tabcolsep{3.0pt}
\tiny
\centering

  \caption{Results of equivalent widths.}
\label{tab:eq}%
    \begin{tabular}{@{}lccccccccccl}
   \tableline\tableline
    aorkey & object & R.A.(degree) & Decl.(degree) & $W_{\rm eq}(10)$  & $W_{\rm eq}(18)$  & $W_{\rm eq}(23)$ & $W_{\rm eq}(28)$ & $W_{\rm eq}(33)$ \\
     \tableline
    10665216 & RAW 631                   & 12.68487  & -72.62765  & 0.33  & 0.68  & 0.42  & 0.78  & 0.46  \\
    10959616 & HD 268835                 & 74.19611  & -69.84022  & 0.08  & 0.15  & 0.10  & 0.13  & 0.11  \\
    12683008 & 2MASS J21380350+5741349   & 324.51465  & 57.69313  & 0.36  & 1.68  & 0.61  & 0.66  & 0.97  \\
    12690688 & HJM C 2-5                 & 167.73296  & -76.75901  & 0.06  & 0.26  & 0.20  & 0.37  & 0.31  \\
    12691969 & V* VY Cha                 & 167.22729  & -77.03692  & 0.04  & 0.21  & 0.19  & 0.23  & 0.12  \\

    \tableline
    \end{tabular}%
 \tablenotetext{a}{Table \ref{tab:eq} is published in its entirety in the
electronic edition of the {\it Astronomical Journal}.  A portion is
shown here for guidance regarding its form and content.}
\tablenotetext{b}{The $W_{\rm eq}(10)$, $W_{\rm eq}(18)$, $W_{\rm eq}(23)$, $W_{\rm eq}(28)$ and $W_{\rm eq}(33)$ are the equivalent widths of the 10, 18, 23, 28 and 33$\mu$m features, respectively.}
\end{table}%

\clearpage

\begin{table}
\setlength\tabcolsep{3.0pt}
\tiny
\centering

  \caption{Results of significance level and crystallinity.}
\label{tab:significanceLevel}%
    \begin{tabular}{@{}lccccccccccl}
   \tableline\tableline
    aorkey & object & R.A.(degree) & Decl.(degree) & $S/U(10)$  & $S/U(18)$  & $S/U(23)$ & $S/U(28)$ & $S/U(33)$ & Crystallinity(\%) & Crystallinity\_min(\%)\\
     \tableline
   10665216 & RAW 631                   & 12.68487 & -72.62765 & 0.53  & 0.45  & 0.65  & 0.85  & 0.56  & 38.5  & 13.1  \\
   10959616 & HD 268835                 & 74.19611 & -69.84022 & 0.19  & 0.14  & 0.18  & 0.15  & 0.14  & 28.0  & 4.5  \\
   12683008 & 2MASS J21380350+5741349   & 324.51465 & 57.69313 & 0.49  & 0.92  & 0.96  & 0.58  & 0.44  & 66.5  & 20.5  \\
   12690688 & HJM C 2-5                 & 167.73296 & -76.75901 & 0.14  & 0.25  & 0.39  & 0.41  & 0.40  & 52.3  & 9.7  \\
   12691969 & V* VY Cha                 & 167.22729 & -77.03692 & 0.09  & 0.17  & 0.35  & 0.25  & 0.16  & 34.0  & 6.6  \\

    \tableline
    \end{tabular}%
 \tablenotetext{a}{Table \ref{tab:significanceLevel} is published in its entirety in the
electronic edition of the {\it Astronomical Journal}.  A portion is
shown here for guidance regarding its form and content.}
\tablenotetext{b}{The $S/U(10)$, $S/U(18)$, $S/U(23)$, $S/U(28)$ and $S/U(33)$ are the significance level of the 10, 18, 23, 28 and 33$\mu$m features of crystalline silicate, respectively.}
\end{table}%

\clearpage

\begin{table}
\setlength\tabcolsep{3.0pt}
\tiny
\centering

  \caption{Results of SIMBAD type.}
 \label{tab:2}%
    \begin{tabular}{@{}lccccccccccl}
   \tableline\tableline
    aorkey & object & R.A.(degree)& Decl.(degree) &SIMBAD type & SIMBAD spectral type \\
     \tableline
    10665216 & RAW 631                   & 12.68487 & -72.62765 & Carbon Star                    & C               \\
    10959616 & HD 268835                 & 74.19611 & -69.84022 & Blue supergiant star & B8[e]           \\
    12683008 & 2MASS J21380350+5741349   & 324.51465 & 57.69313 & T Tau-type Star & G9              \\
    12690688 & HJM C 2-5                 & 167.73296 & -76.75901 & Pre-main sequence Star & M6e             \\
    12691969 & V* VY Cha                 & 167.22729 & -77.03692 & Variable Star of Orion Type & M0.5            \\
    \tableline
    \end{tabular}%
 \tablenotetext{a}{Table \ref{tab:2} is published in its entirety in the
electronic edition of the {\it Astronomical Journal}.  A portion is
shown here for guidance regarding its form and content.}
\end{table}%

\clearpage
\begin{table}
\setlength\tabcolsep{1.0pt}
\tiny

\centering
  \caption{Summary of SIMBAD type. The number is the amount of corresponding SIMBAD type.}
 \label{tab:3}%
    \begin{tabular}{cccccc}
    \tableline\tableline
    SIMBAD type & number & SIMBAD type & number & SIMBAD type & number\\
      \tableline
Young Stellar Object & 132   & Planetary Nebula & 8     & Galaxy & 1 \\
Variable Star of Orion Type & 128   & Brown Dwarf (M<0.08solMass) & 7     & Reflection Nebula & 1 \\
T Tau-type Star & 100   & Variable Star of RV Tau type & 6     & Possible Planetary Nebula & 1 \\
Herbig Ae/Be star & 55    & Blue supergiant star & 5     & Variable of BY Dra type & 1 \\
Pre-main sequence Star & 54    & Flare Star & 5     & Dark Cloud (nebula) & 1 \\
Emission-line Star & 36    & Double or multiple star & 4     & Star in Association & 1 \\
Star  & 28    & Nova  & 3     & Variable Star of irregular type & 1 \\
Variable Star of Mira Cet type & 25    & S Star & 3     & Wolf-Rayet Star & 1 \\
Post-AGB Star (proto-PN) & 20    & Seyfert 1 Galaxy & 3     & Possible Red supergiant star & 1 \\
Red supergiant star & 19    & Low-mass star (M<1solMass) & 2     & Star suspected of Variability & 1 \\
Variable Star & 18    & Emission Object & 2     & Variable Star of RR Lyr type & 1 \\
Infra-Red source & 15    & Variable Star of FU Ori type & 2     & Extra-solar Confirmed Planet & 1 \\
Young Stellar Object Candidate & 14    & X-ray source & 2     & Variable Star of delta Sct type & 1 \\
T Tau star Candidate & 12    & Dense core & 2     & Variable Star with rapid variations & 1 \\
Star in Cluster & 12    & Evolved supergiant star & 2     & SuperNova Remnant & 1 \\
Asymptotic Giant Branch Star (He-burning) & 12    & Be Star & 2     & Pulsating variable Star & 1 \\
OH/IR star & 11    & Carbon Star & 1     & Brown Dwarf Candidate & 1 \\
Long-period variable star & 11    & Object of unknown nature & 1     &       &  \\
Semi-regular pulsating Star & 9     & Post-AGB Star Candidate & 1     &       &  \\

    \tableline
    \end{tabular}%

\end{table}%

\clearpage
\begin{table}
\setlength\tabcolsep{1.5pt}
\tiny
\centering
    \caption{Results of cross-identification with LAMOST.}
\label{tab:4}%
    \begin{tabular}{@{}lccccccccc}
    \tableline\tableline
    aorkey  & object & R.A.(degree) & Decl.(degree) & class & subclass & z & H$\alpha$ & SIMBAD type & SIMBAD spectra type\\
    \tableline
    14971648 & 2MASS J03022104+1710342   & 45.58757 & 17.17621 & STAR  & M3    & -2.59E-04 & 1     & T Tau-type Star & M3              \\
    5633280 & NAME LDN 1455 IRS 2       & 51.94873 & 30.20122 & STAR  & K5    & -7.74E-04 & 1     & Dense core             & ~               \\
    22031616 & EM* LkHA 352A             & 52.2125 & 31.3051 & STAR  & M5    & 3.32E-05 & 1     & Emission-line Star     & M4.5            \\
    22032384 & 2MASS J03279019+3119548   & 52.21324 & 31.33186 & STAR  & M0    & 5.19E-05 & 1     & Young Stellar Object & K7.0            \\
    22032128 & 2MASS J03285663+3118356   & 52.23595 & 31.30986 & STAR  & M2    & 2.12E-03 & 1     & Young Stellar Object & M1.5            \\

    \tableline
    \end{tabular}%
  \tablenotetext{a}{Table \ref{tab:4} is published in its entirety in the
electronic edition of the {\it Astronomical Journal}.  A portion is
shown here for guidance regarding its form and content.}
\tablenotetext{b}{z is the redshift of object. In the column of H$\alpha$, 1, 0 and -1 mean the object has a strong H$\alpha$ emission line, no strong H$\alpha$ line and a strong H$\alpha$ absorption line, respectively.}

  \medskip

\end{table}%

\clearpage
\begin{table}

\setlength\tabcolsep{3.0pt}
\centering

  \caption{LAMOST spectra type summary. The number is the amount of targets.}
 \label{tab:5}%
    \begin{tabular}{@{}ccccc}
   \tableline\tableline

    spectra type & number & number(H$\alpha$=1) & number(H$\alpha$=0) & number(H$\alpha$=-1)\\

     \tableline
   M-type star & 56 & 54 & 2 & 0 \\
   K-type star & 15 & 15 & 0 & 0 \\
   G-type star & 12 & 7 & 4 & 1 \\
   F-type star & 3 & 3 & 0 & 0 \\
   A-type star & 3 & 3 & 0 & 0 \\
   White Dwarf & 1 & 1 & 0 & 0 \\
   QSO & 1 & 0 & 1 & 0 \\

    \tableline
    \end{tabular}%

\end{table}%

\end{document}